\documentclass[aps,prd,groupedaddress,showpacs,nofootinbib,amssymb,balancelastpage,preprintnumbers,floatfix]{revtex4}

\usepackage[dvipdfmx]{graphicx}
\usepackage{graphicx,bm,color}

\usepackage{amsmath}
\usepackage{amssymb}
\usepackage{amsfonts}
\usepackage{cases}
\usepackage{cancel}
\usepackage[normalem]{ulem}
\usepackage{subfigure}
\usepackage{here}
\usepackage{color}
\usepackage{comment}
\usepackage{hyperref}

\usepackage{tikz}
\usetikzlibrary{matrix}

\allowdisplaybreaks[3]

\newcommand{\cred}[1]{{\color{black}#1}}

\begin{document}
	

\title{First-order CP phase transition in two-flavor QCD  
at $\theta = \pi$ under electromagnetic scale anomaly via a Nambu-Jona-Lasinio description}

\author{Yuanyuan Wang}\thanks{{\tt yuanyuanw23@jlu.edu.cn}}
\affiliation{Center for Theoretical Physics and College of Physics, Jilin University, Changchun, 130012,
China}

\author{Mamiya Kawaguchi}\thanks{{\tt mamiya@aust.edu.cn}} 
      \affiliation{ 
Center for Fundamental Physics, School of Mechanics and Physics,
Anhui University of Science and Technology, Huainan, Anhui 232001, People’s Republic of China
}

\author{Shinya Matsuzaki}\thanks{{\tt synya@jlu.edu.cn}}
\affiliation{Center for Theoretical Physics and College of Physics, Jilin University, Changchun, 130012, China}%

\author{Akio Tomiya}\thanks{{\tt akio@yukawa.kyoto-u.ac.jp}} 
\affiliation{Department of Information and Mathematical Sciences, Tokyo Woman’s Christian  University, Tokyo 167-8585, Japan} 
\affiliation{RIKEN Center for Computational Science, Kobe 650-0047, Japan} 

\begin{abstract}
We discuss the thermal CP phase transition in QCD at $\theta=\pi$ under a weak magnetic field background, where the electromagnetic scale anomaly gets significant. 
To explicitize, we work on a two-flavor \cred{Nambu-Jona-Lasinio model} at $\theta=\pi$ in the mean field approximation, including the electromagnetic-scale anomaly term. 
We find that the thermal CP phase transition becomes first order 
and the strength of the first order gets more prominent as the magnetic field increases. 
The associated potential barrier is thermally created by the electromagnetic scale
anomaly and gives rise to criticality due to the induced potential of a non-perturbative form $\sim \frac{|eB|^3}{f_\pi} \frac{|P|}{P^2 + m_0^2}$,
where $eB$ denotes the magnetic field strength;  
$P$ the CP order parameter, and $m_0$ the isospin-symmetric current-quark mass. 
\end{abstract}

\maketitle

\section{Introduction}

Exploring CP violation is one of great importance to comprehend the cosmic history of the universe; e.g., pursuit of the origin of matter and anti-matter. 
This issue is essentially linked to a deep understanding of  
the topological feature of the non-Abelian gauge theories. 
In the Standard Model of particle physics, 
the so-called topological $\theta$ term in quantum chromodynamics (QCD) 
violates the CP symmetry 
unless $\theta=0$ or $\pi$. 
To the size of this $\theta$, the neutron electric dipole moment measurements have so far suggested $\theta$  
to be $< 10^{-11}$~\cite{Abel:2020pzs,Liang:2023jfj}. 
Thus the present-day strong CP violation has to be extremely small somehow with $\theta \sim 0$, not allowing for $\theta =\pi$.

The case might be quite different in the early hot universe: 
a local CP-odd domain may be created in hot QCD plasma 
due to the presence of the QCD sphaleron~\cite{Manton:1983nd,Klinkhamer:1984di}, so that the QCD vacuum characterized by 
the strong CP phase $\theta$ and its fluctuation (in the spatial-homogeneous direction) gets significantly sizable~\cite{Kharzeev:2007tn,Kharzeev:2007jp,Fukushima:2008xe} within the QCD time scale~\cite{McLerran:1990de,Moore:1997im,Moore:1999fs,Bodeker:1999gx}. 
Given this feasibility, QCD at $\theta=\pi$ is still well motivated to investigate 
in a viewpoint of not only the purely theoretical interest, but also even phenomenology 
or cosmology in the early universe.

The dynamics of the CP symmetry in Yang-Mills theories with or without quarks at $\theta=\pi$ has extensively been discussed so far. 
One of the key ``benchmark"s is the so-called Dashen's phenomenon~\cite{Dashen:1970et},  
that suggested the spontaneous breaking of the CP symmetry at $\theta=\pi$ based 
on a chiral effective model approach. 
In earlier dates, instanton gas models and deconfinement models also developed this phenomenon to finite temperature, which involves
the nature of the deconfinement phase transition~\cite{Gross:1980br,Weiss:1980rj}: 
the CP symmetry and/or the center symmetry is spontaneously broken at lower temperatures, but is manifest at higher temperatures.    
This feature has been well captured in the large-$N_c$ limit of QCD with quarks~\cite{Witten:1980sp,Witten:1998uka}, where
the CP symmetry at $\theta=\pi$ is predicted to be spontaneously broken at low temperatures. 
Thus, CP symmetry $\theta=\pi$ is thought of also as a crucial ingredient in understanding the confinement.

Recently, the higher-form symmetry~\cite{Gaiotto:2014kfa} incorporated into the 't Hooft anomaly matching~\cite{tHooft:1979rat,Frishman:1980dq} has developed the study along this research line. It has been predicted that at $\theta=\pi$, 
either the CP or the center symmetry should be spontaneously broken unless the theory becomes gapless (i.e., deconfined):  
the theory cannot be in a gapped confined phase without breaking the
CP symmetry~\cite{Gaiotto:2017yup,Cordova:2019bsd}, namely, 
$T_c^{\rm (CP)} > T_c^{\rm (dec)}$.

For nonzero $\theta$, lattice simulations under the standard base of the 
MonteCarlo method unfortunately suffers from infamous sign problem. To avoid the sign problem, interesting methods for nonzero $\theta$ have so far been proposed. 
In particular, the so-called subvolume method has been applied to four-dimensional $SU(2)$ Yang-Mills theory, which suggests that the CP symmetry at $\theta=\pi$ is broken at zero temperature and is restored at higher temperatures~\cite{Kitano:2020mfk,Kitano:2021jho,Yamada:2024pjy,Yamada:2024vsk}. 
Furthermore, very recently, another technique based on the analytic continuation 
from the imaginary $\theta$ to real $\theta$ has been applied to the lattice simulation 
and also observed the same phenomenon~\cite{Hirasawa:2024fjt}.

In a realistic setup including quarks, where $\theta$ is transformed by the axial rotation to be able to have dependence on the number of quark flavors $N_f$ as $\theta \to \theta_{N_f}=\theta/N_f$ with identical masses, 
the analytic continuation method can still be 
utilized to get expectation values for physical observables~\cite{DelDebbio:2002xa,DElia:2003zne,DelDebbio:2006sbu,Giusti:2007tu,Izubuchi:2007rmy,Vicari:2008jw,DElia:2013uaf,Hirasawa:2024vns}. 
However, those calculations have been done away from the physical point.
The subvolume method proposed in~\cite{Kitano:2021jho} is also applicable even in the case with quarks, but it is still hard to reach physical results. 
Another idea based on digital and analog quantum simulations and calculations has been adopted to the case with nonzero $\theta$ using tensor networks, which 
have been applied only for toy models, but not for the realistic setup like four dimensional QCD at the physical point~\cite{Byrnes:2002gj,Funcke:2019zna,Kuramashi:2019cgs,Chakraborty:2020uhf,Honda:2021aum,Honda:2022hyu}.
Thus, reaching realistic QCD with quarks is still challenging for lattice simulations, so the spontaneous CP symmetry breaking and its restoration still involves lots of unknowns.

On the other hand, QCD at $\theta=\pi$ has so far been explored 
also by various chiral effective models~\cite{Dashen:1970et,Witten:1980sp,Pisarski:1996ne,Creutz:2003xu,Mizher:2008hf,Boer:2008ct,Creutz:2009kx,Boomsma:2009eh,Sakai:2011gs,Chatterjee:2011yz,Sasaki:2011cj,Sasaki:2013ewa,Aoki:2014moa,Mameda:2014cxa,Verbaarschot:2014upa,Bai:2023cqj,Huang:2024nbd}. 
The field of target physics is nowadays getting broader, which has been originated from the Dashen phenomenon up until the recent interest in gravitational wave productions based on the expected CP phase transition of the first-order or the domain wall collapse of the axionlike particle at $\theta=\pi$. 
Those studies have provided benchmarks toward understanding of 
QCD at $\theta=\pi$, prior to the future development in lattice simulations. 

In light of the early universe (and/or heavy ion collision experiments), 
however, the CP phase transition might be more involved: 
a strong electromagnetic field could be generated via cosmological first-order phase transitions~\cite{Vachaspati:1991nm,Enqvist:1993np,Grasso:1997nx,Grasso:2000wj,Ellis:2019tjf,Zhang:2019vsb,Di:2020kbw,Yang:2021uid,Saga:2023afb,Liu:2024mdo,Di:2024gsl}, 
or inflationary scenarios~\cite{Turner:1987vd,Garretson:1992vt,Anber:2006xt,Domcke:2019mnd,Domcke:2019qmm,Patel:2019isj,Domcke:2020zez,Shtanov:2020gjp,Okano:2020uyr,Cado:2021bia,Kushwaha:2021csq,Gorbar:2021rlt,Gorbar:2021zlr,Gorbar:2021ajq,Fujita:2022fwc,Cado:2022pxk,Domcke:2022kfs,Cado:2023gan,Murai:2023gkv,Bastero-Gil:2023htv,Bastero-Gil:2023mxm,Otsuka:2024mdg,vonEckardstein:2024tix,Sharma:2024nfu}. 
The strength of the produced electromagnetic field would be redshifted to be smaller and smaller toward a later time epoch including the timing when the universe undergoes the QCD phase transition.  
With the active QCD sphaleron taken into account~\cite{McLerran:1990de,Moore:1997im,Moore:1999fs,Bodeker:1999gx}, 
thereby, the CP phase transition at $\theta =\pi$ can be under 
such a redshifed background electromagnetic field. 

Although the electromagnetic field does not break the CP symmetry additionally, 
it does explicitly break the chiral and $U(1)$ axial symmetries because of  
the different charges for quarks. 
This breaking can cause the so-called magnetic catalysis or inverse magnetic catalysis~\cite{Bali:2011qj}, where the constituent dynamical quark masses and 
the quark condensates are amplified or suppressed by the applied magnetic field. 
At $\theta = \pi$ (i.e., $\theta_{N_f}=\pi/N_f$), the scalar quark condensate will be interchanged with 
the pseudoscalar quark condensate via the axial rotation, where the latter plays the role of the order parameter for the CP symmetry breaking as well as the chiral order parameter. 
Therefore, the CP symmetry breaking in the same way as in the chiral symmetry breaking is expected to be enhanced or suppressed in the presence of a magnetic field. 
In fact, this kind of phenomenon has been observed in the chiral effective model approaches~\cite{Chatterjee:2014csa,Bandyopadhyay:2019pml,Carlomagno:2025ayh}.

In a view of a cosmic magnetic field background at the QCD epoch as aforementioned, 
the magnetic field strength might be redshifted to be as small as or smaller than \cred{the intrinsic scale of QCD, $\Lambda_{\rm QCD} \sim 200 - 400$ MeV:  
$eB \lesssim \Lambda_{\rm QCD}^2$}. 
In that case, it has recently been clarified that the electromagnetic-scale anomaly effect is not negligible in a weak magnetic field and is shown to have important implications for the chiral phase transition properties regarding the quark mass dependence in QCD with $\theta=0$~\cite{Kawaguchi:2021nsa,YuanyuanWang:2022nds,YuanyuanWang:2024zwf}. 
This scale anomaly effect has never been examined along with nonzero $\theta$.

In this paper, we discuss the effect of the electromagnetic 
scale anomaly by explicitly working on a two-flavor 
\cred{NJL model} at $\theta=\pi$ in the mean field approximation (MFA). 
We observe that the thermomagnetic correction part in the electromagnetic scale anomaly term develops a potential barrier around the origin of the direction of the CP order parameter. 
The associated potential barrier is thermally created by the electromagnetic scale
anomaly and gives rise to criticality due to the induced potential of a nonperturbative form, 
 which cannot be described by the Ginzburg-Landau description. $\sim \frac{|eB|^3}{f_\pi} \frac{|P|}{P^2 + m_0^2}$, 
where $eB$ denotes the magnetic field strength;  
$P$ the CP order parameter, and $m_0$ the isospin-symmetric current-quark mass.


\section{ NJL model with electromagnetic scale anomaly}

We begin by introducing the two-flavor \cred{NJL} model with nonzero $\theta$ 
under a constant external magnetic field. 
The Lagrangian of this model is as follows:
\begin{equation}
\begin{aligned}
\mathcal{L}= & \bar{\psi}\left(i \gamma^\mu D_\mu- m_0 \cdot {\bf 1}_{ 2 \times 2}\right) \psi+\frac{g_s}{2} \sum_{a=0}^3\left[\left(\bar{\psi} \tau_a \psi\right)^2+\left(\bar{\psi} i \gamma_5 \tau_a \psi\right)^2\right] \\
& +g_d\left\{e^{i \theta} \operatorname{det}\left[\bar{\psi}\left(1+\gamma_5\right) \psi\right]+e^{-i \theta} \operatorname{det}\left[\bar{\psi}\left(1-\gamma_5\right) \psi\right]\right\}
\end{aligned}
\label{Lag:PNJL}
\end{equation}
where $\psi=(u, d)^T$ denotes the two-flavor quark field and the isospin-symmetric current-quark mass $m_0$. 
The covariant derivative \cred{$D_\mu=\partial_\mu-i q A^{EM}_\mu$ contains the couplings of quarks to} 
the external electromagnetic field $A^{EM}_\mu$, with the electromagnetic charge matrix given by 
 $q=e \cdot \operatorname{diag}\{2 /3,-1 /3\}$.  
For simplicity, the magnetic field $B$ is set along the $z$-axis and is embedded in the electromagnetic gauge field of the form $A^{EM}_\mu=(0,0, B x, 0)$. 
The four-fermion interaction with the coupling $g_s$ keeps the full chiral $U(2)_L \times U(2)_R$ invariance, 
 and $\tau_a(a=0, \ldots,3)$ denotes the $U(2)$ generators including $\tau_0 = {\bf 1}_{2 \times 2}$. 
 The term with the coupling $g_d$ is called the Kobayashi-Maskawa-’t Hooft (KMT) determinant term~\cite{Kobayashi:1970ji,Kobayashi:1971qz,tHooft:1976rip,tHooft:1976snw} and preserves the $S U(2)_L \times S U(2)_R \times U(1)_V$ invariance, 
 but breaks the $U(1)$ axial symmetry.

 We introduce the three-dimensional momentum cutoff $\Lambda$ to regularize the \cred{NJL} model in the MFA. 
We refer to empirical hadron observables in the isospin symmetric limit 
at $e B=T=0$~\cite{Boer:2008ct,Boomsma:2009eh,Sakai:2011gs}: the pion mass $m_\pi=0.1402\, \mathrm{GeV}$, the pion decay constant $f_\pi=0.0926\, \mathrm{GeV}$, and quark condensate $\langle\bar{u} u\rangle=\langle\bar{d} d\rangle=(-0.2415 \,  \mathrm{GeV})^3$. 
Then the model parameters can be determined as $m=0.006 \, \mathrm{GeV}$, $\Lambda=0.59 \, \mathrm{GeV}$, and for $g_s=2(1-c) G_0$ with $g_d=2 c G_0$ and 
$c=0.2$~\cite{Boer:2008ct,Boomsma:2009eh,Sakai:2011gs}, we have $G_0 \Lambda^2=2.435$.

We utilize the imaginary time formalism and the Landau level decomposition to evaluate the thermomagnetic contributions to the quark condensate taking the following replacements:
\begin{equation}
\begin{aligned}
p_0 & \leftrightarrow i \omega_{\mathbf{k}}=i(2 \mathbf{k}+1) \pi T \\
\int \frac{d^4 p}{(2 \pi)^4} & \leftrightarrow i T \sum_{\mathbf{k}=-\infty}^{\infty} \int \frac{d^3 p}{(2 \pi)^3} \\
& \leftrightarrow i T \sum_{\mathbf{k}=-\infty}^{\infty} \sum_{n=0}^{\infty} \alpha_n \frac{\left|q_f B\right|}{4 \pi} \int_{-\infty}^{\infty} \frac{d p_z}{2 \pi} f_{\Lambda}\left(p_z, n\right).
\end{aligned}
\label{replacement}
\end{equation}
$\omega_k$ represents the Matsubara frequency with $k$ as an integer, while $n$ denotes the Landau levels. The factor  
 $\alpha_n=2-\delta_{n, 0}$ refers to the spin degeneracy for the Landau levels. 
 By applying these replacements, the chiral phase transition can be directly extended from the vacuum to the case of $T \neq 0$ and $e B \neq 0$. In addition we introduce a soft-cutoff regulator function~\cite{Frasca:2011zn}, 
 \begin{equation}
f_{\Lambda}\left(p_z, n\right)=\frac{\Lambda^{10}}{\Lambda^{10}+\left(p_z^2+2 n\left|q_f B\right|\right)^5} 
\,. 
\label{reg}
\end{equation}

It is convenient to transfer the $\theta$ dependence attached on the $g_d$ term in Eq.(\ref{Lag:PNJL})  
from the $g_d$ term to the mass term by a $U(1)$ axial rotation of $\psi$ as 
$\psi \rightarrow e^{-i \gamma_5 \frac{\theta}{4}} \psi \equiv \psi^{\prime}$. 
Then the Lagrangian in Eq.(\ref{Lag:PNJL}) can be rewritten as
 \begin{equation}
\begin{aligned}
\mathcal{L} \rightarrow \mathcal{L}^{\prime}= & \bar{\psi^{\prime}}\left(i \gamma^\mu D_\mu-\mathbf{m}(\theta)\right) \psi^{\prime}+\frac{g_s}{2} \sum_{a=0}^3\left[\left(\bar{\psi^{\prime}} \tau_a \psi^{\prime}\right)^2+\left(\bar{\psi^{\prime}} i \gamma_5 \tau_a \psi^{\prime}\right)^2\right] \\
& +g_d\left\{ \operatorname{det}\left[\bar{\psi^{\prime}}\left(1+\gamma_5\right) \psi^{\prime}\right]+ \operatorname{det}\left[\bar{\psi^{\prime}}\left(1-\gamma_5\right) \psi^{\prime}\right]\right\} 
\end{aligned}
\label{Lag:PNJL:prime}
\end{equation}
where 
\begin{align}
\mathbf{m}(\theta)
&=m_0 \left[\cos \frac{\theta}{2}+i \gamma_5 \sin \frac{\theta}{2}\right] 
\,, \notag\\ 
& =m_0 \left[\cos \theta_{N_f=2}+i \gamma_5 \sin \theta_{N_f=2} \right] 
\,, 
\end{align}
and all the $\theta$ dependence has been now transformed into the complex mass $\mathbf{m}(\theta)$. The scalar and pseudoscalar bilinears in the original base are related to those in the prime base:
\begin{equation}
\begin{aligned}
\left(\bar{\psi} \psi\right) & =\left(\bar{\psi}^{\prime} \psi^{\prime}\right) \cos \frac{\theta}{2}+\left(\bar{\psi}^{\prime} i \gamma_5 \psi_i^{\prime}\right) \sin \frac{\theta}{2} \,, \\
\left(\bar{\psi} i \gamma_5 \psi\right) & =-\left(\bar{\psi}^{\prime} \psi^{\prime}\right) \sin \frac{\theta}{2}+\left(\bar{\psi}^{\prime} i \gamma_5 \psi^{\prime}\right) \cos \frac{\theta}{2} .
\end{aligned}
\end{equation}

We work in the MFA so that 
the isospin-singlet scalar-quark bilinear $\bar{\psi} \psi=\bar{u} u+\bar{d} d$ and its CP partner $\bar{\psi} i \gamma_5 \psi=\bar{u}i \gamma_5 u+\bar{d} i \gamma_5d$ 
can be expanded around the mean fields $S_{q}^{\prime}=\left\langle\bar{q}^{\prime} q^{\prime}\right\rangle(q=u,d)$, and $P_{q}^{\prime}=\left\langle\bar{q}^{\prime}i \gamma_5 q^{\prime}\right\rangle(q=u,d)$. 
The MF values for other isotriplet quark-bilinears are set to zero. 
Thus, in the MFA, the Lagrangian in Eq.(\ref{Lag:PNJL:prime}) takes the form
\begin{equation} 
\begin{aligned}
\mathcal{L}_{\mathrm{MFA}}=& \bar{\psi}^{\prime}\left(i \gamma^\mu D_\mu-\mathbf{M}(\theta,S_{u}^{\prime},S_{d}^{\prime},P_{u}^{\prime},P_{d}^{\prime})\right) \psi^{\prime}-\frac{g_s}{2} \left[\left(S_{u}^{\prime}+S_{d}^{\prime}\right)^2+\left(P_{u}^{\prime}+P_{d}^{\prime}\right)^2\right] \\
& -2g_d\left(S_{u}^{\prime}S_{d}^{\prime}-P_{u}^{\prime}P_{d}^{\prime}\right)
\end{aligned}
\end{equation}
where 
\begin{align} 
\mathbf{M}(\theta,S_{u}^{\prime},S_{d}^{\prime},P_{u}^{\prime},P_{d}^{\prime})=\operatorname{diag}\{M_{u}(\theta,S_{u}^{\prime},S_{d}^{\prime},P_{u}^{\prime},P_{d}^{\prime}),  M_{d}(\theta,S_{u}^{\prime},S_{d}^{\prime},P_{u}^{\prime},P_{d}^{\prime})\} 
\,, 
\end{align} 
and 
\begin{equation}
\begin{aligned}
M_{u}(\theta,S_{u}^{\prime},S_{d}^{\prime},P_{u}^{\prime},P_{d}^{\prime})& =\alpha_{u}\left(\theta, S_{u}^{\prime},S_{d}^{\prime} \right)+i \gamma_5 \beta_{u}\left(\theta,P_{u}^{\prime},P_{d}^{\prime}\right) 
\,, \\
\alpha_{u}\left(\theta, S_{u}^{\prime},S_{d}^{\prime} \right) 
&=m_0 \cos \frac{\theta}{2}-g_s\left(S_{u}^{\prime}+S_{d}^{\prime}\right)-2g_d S_{d}^{\prime}, \\ 
& \equiv m_0 \cos \frac{\theta}{2} + \sigma_u' 
\,, \\ 
\beta_{u}\left(\theta,P_{u}^{\prime},P_{d}^{\prime}\right) 
& =m_0 \sin \frac{\theta}{2}-g_s\left(P_{u}^{\prime}+P_{d}^{\prime}\right)+2g_d P_{d}^{\prime}, 
\,, \\ 
& \equiv m_0 \sin \frac{\theta}{2} + \eta_u' 
\,, \\ 
M_{d}(\theta,S_{u}^{\prime},S_{d}^{\prime},P_{u}^{\prime},P_{d}^{\prime})& =\alpha_{d}\left(\theta, S_{u}^{\prime},S_{d}^{\prime} \right)+i \gamma_5 \beta_{d}\left(\theta,P_{u}^{\prime},P_{d}^{\prime}\right), \\
\alpha_{d}\left(\theta, S_{u}^{\prime},S_{d}^{\prime} \right) 
& =m_0 \cos \frac{\theta}{2}-g_s\left(S_{u}^{\prime}+S_{d}^{\prime}\right)-2g_d S_{u}^{\prime}, \\ 
& = m_0 \cos \frac{\theta}{2} + \sigma_d' 
\,, \\ 
\beta_{d}\left(\theta,P_{u}^{\prime},P_{d}^{\prime}\right) 
& =m_0 \sin \frac{\theta}{2}-g_s\left(P_{u}^{\prime}+P_{d}^{\prime}\right)+2g_d P_{u}^{\prime},, \\ 
& \equiv m_0 \sin \frac{\theta}{2} + \eta_d' 
\,. 
\end{aligned}
\label{alpha-beta}
\end{equation}
The thermomagnetic potential in the MFA is obtained by integrating out the quarks as follows:
\begin{equation}
\begin{aligned}
\Omega\left(\theta, S_u^{\prime}, S_d^{\prime}, P_u^{\prime}, P_d^{\prime}, 
T, e B\right) & =\frac{g_s}{2}\left[\left(S_u^{\prime}+S_d^{\prime}\right)^2+\left(P_u^{\prime}+P_d^{\prime}\right)^2\right]+2 g_d\left(S_u^{\prime} S_d^{\prime}-P_u^{\prime} P_d^{\prime}\right)\\&- N_c \sum_{f=u, d}\sum_{n=0}^{\infty} \alpha_n \int \frac{d p_z}{2 \pi} \frac{\left|q_f B\right|}{2 \pi} f_{\Lambda}\left(p_z, n\right)\left[
E_{f, n} 
+2 T \ln \left(1+    e^{-\frac{E_{f, n}}{T}} \right)\right] 
\end{aligned}
\label{Omega:normal}
\end{equation}
where 
\begin{align} 
M^2_f &=\alpha^2_f+\beta^2_f
\,, \notag\\ 
E_{f, n} &=\sqrt{M^2_f+2n \left|q_f B\right|+p^2_z} 
\,,  
\end{align} 
for $f=u$ and $d$.

Besides, we incorporate the electromagnetic scale anomaly into the PNJL model~\cite{Kawaguchi:2021nsa,YuanyuanWang:2022nds,YuanyuanWang:2024zwf}: 
\begin{equation}
V_{\mathrm{eff}}^{(\mathrm{Tad})} 
=-\frac{\varphi}{f_{\varphi}} T_\mu^\mu \,,    
\label{Tad}
\end{equation}
where 
\begin{align} 
\varphi = \sqrt{ \sum_{f=u,d} \left[ (\sigma_f')^2 + (\eta_f')^2 \right]}  
\,,   
\end{align} 
and $f_\varphi$ stands for the vacuum expectation value of this $\varphi$ at $T=eB=0$. 
What we are concerned about is in a weak field limit,  
$e B \ll \cred{\Lambda_{\rm QCD}^2}$, as noted in the Introduction~\footnote{
When $eB$ exceeds the threshold strength $\gtrsim \cred{\Lambda_{\rm QCD}^2}$, 
the electromagnetic scale anomaly gets started to suppress.    
This is due to the fact that above the threshold, 
the dynamics of quarks will be dominated by the lowest Landau level states, which are longitudinally polarized along the magnetic field direction. 
Since the electromagnetic scale anomaly develops along the transverse direction, 
its effect becomes mostly remarkable in the weak magnetic regime~\cite{Kawaguchi:2021nsa}.}.    
Then the trace anomaly $T_\mu^\mu$ in Eq.(\ref{Tad}) takes the form~\cite{YuanyuanWang:2024zwf} 
\begin{equation}
T_\mu^\mu 
=\frac{\beta(e)}{e^3} |e B|^2+\frac{1}{2}N_c \sum_{f=u,d,s}\sum_{n=0}^{\infty} \alpha_n\frac{ q_f^2}{e^2}M^2_f \frac{\left|q_f B\right|}{4 \pi} \int_{-\infty}^{\infty} \frac{d p_z}{2 \pi} F\left(T,e B, M_f\right)|e B|^2. 
\label{tr-anom}
\end{equation}
The first term $\beta(e)$ in Eq.(\ref{tr-anom}) denotes the beta-function coefficient of the electromagnetic coupling $e$ at the one-loop level, which is evaluated as 
\begin{equation}
\beta(e)=\frac{1}{(4 \pi)^2} \frac{4 N_c}{3} \sum_f e\, q_f^2
\,, 
\end{equation}
 and represents the electromagnetic scale anomaly contribution in the magnetized vacuum without $T$. 
 The second term in Eq.(\ref{tr-anom}) is the thermomagnetic part including the Fermi-Dirac thermal distribution function as~\footnote{Higher-order contributions in powers of $eB$ are potentially included in 
the full quark mass $M_f$ in the second term of Eq.(\ref{tr-anom}) , 
while other possible higher order terms in $eB$ have been 
disregarded in Eq.(\ref{tr-anom}). 
This may be a crude truncation. 
In~\cite{Ghosh:2019kmf} a nonperturbative evaluation of the photon polarization function coupled to the chiral singlet meson field has been made attempted. This could it possible to make a straightforward computation of the photon polarization in the weak magnetic field. 
The reliability of the present truncation prescription could be evaluated in such a way, which is to be left in the future work.}   
\begin{equation}
\begin{aligned}
& F\left(T, e B, M_f\right)=\frac{1}{E^5_{f, n}}\cdot\frac{-2}{\exp \left(\frac{E_{f, n}}{T}\right)+1}, \qquad {\rm with} \qquad 
& E^2_{f, n}=p_z^2+2 n\left|q_f B\right|+M_f^2, 
\end{aligned} \label{F} 
\end{equation}

The total thermomagnetic potential in the MFA at nonzero $\theta$ containing the Polyakov loop potential and the tadpole potential is thus given as 
the sum of two contributions in Eqs.(\ref{Omega:normal}) and (\ref{Tad}) with 
Eq.(\ref{tr-anom}), 
\begin{equation}
\Omega_{tot}\left(\theta, S_u^{\prime}, S_d^{\prime}, P_u^{\prime}, P_d^{\prime}, 
T, e B\right)=\Omega\left(\theta, S_u^{\prime}, S_d^{\prime}, P_u^{\prime}, P_d^{\prime}, 
T, e B\right)+V_{\text {eff }}^{(\mathrm{Tad})}\left(\theta, S_u^{\prime}, S_d^{\prime}, P_u^{\prime}, P_d^{\prime}, T, e B\right). \label{Omega-tot}
\end{equation}
When $T$ and $e B$ are given with $\theta = \pi$, 
the order parameters of the \cred{chiral and CP phase transitions} can be determined by the following stationary conditions: 
\begin{equation}
\begin{aligned}
&\frac{\partial \Omega_{tot}\left(\theta=\pi, S_u^{\prime}, S_d^{\prime}, P_u^{\prime}, P_d^{\prime}, 
T, e B\right)}{\partial S_f^{\prime}}=0,\quad  \frac{\partial \Omega_{tot}\left(\theta=\pi, S_u^{\prime}, S_d^{\prime}, P_u^{\prime}, P_d^{\prime}, 
T, e B\right)}{\partial P_f^{\prime}}=0,\\
\end{aligned}
\end{equation}

\section{CP phase transition at $\theta=\pi$}

\begin{figure}[t] 
\centering 
\includegraphics[width=0.49\textwidth]{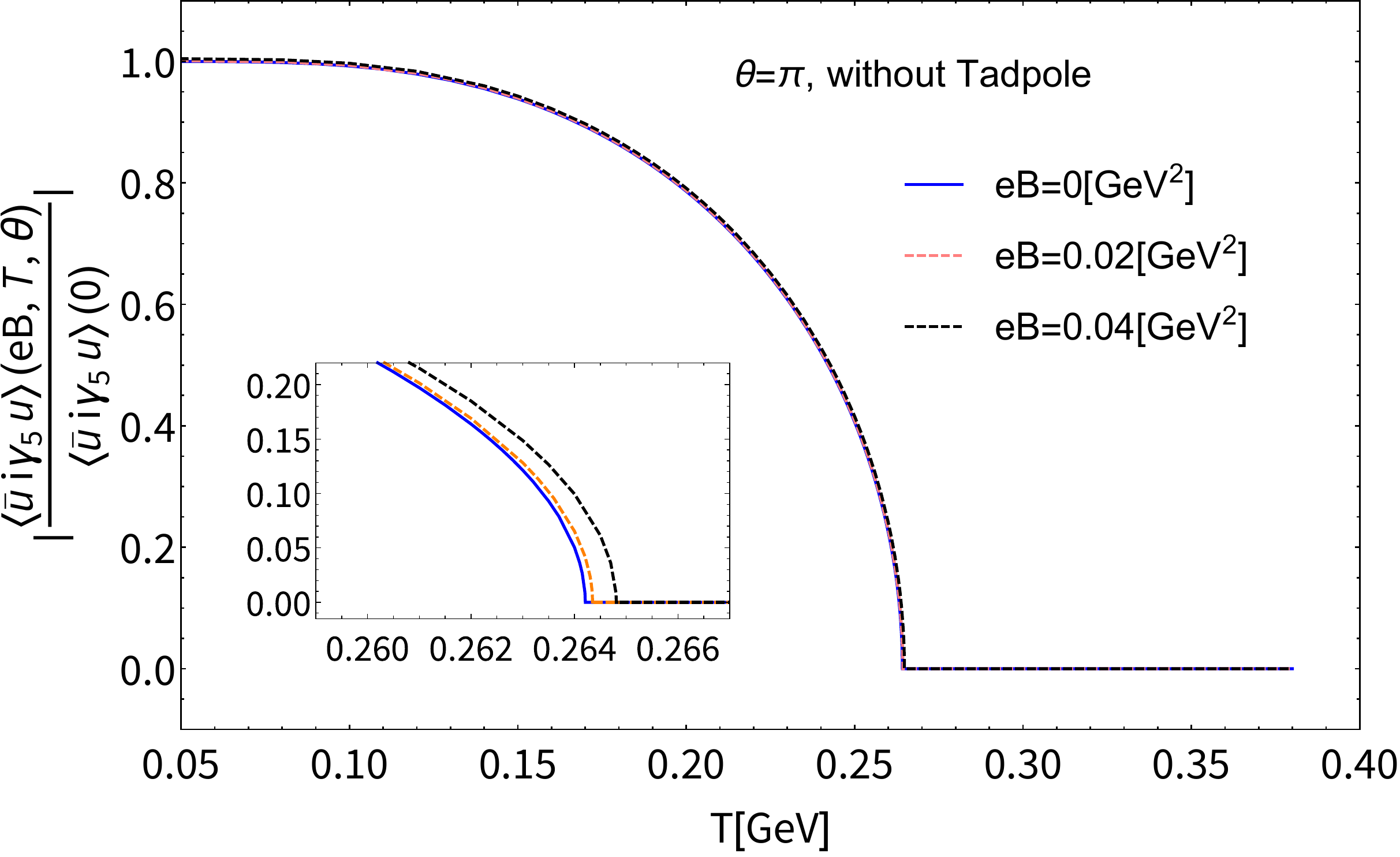}
\includegraphics[width=0.49\textwidth]{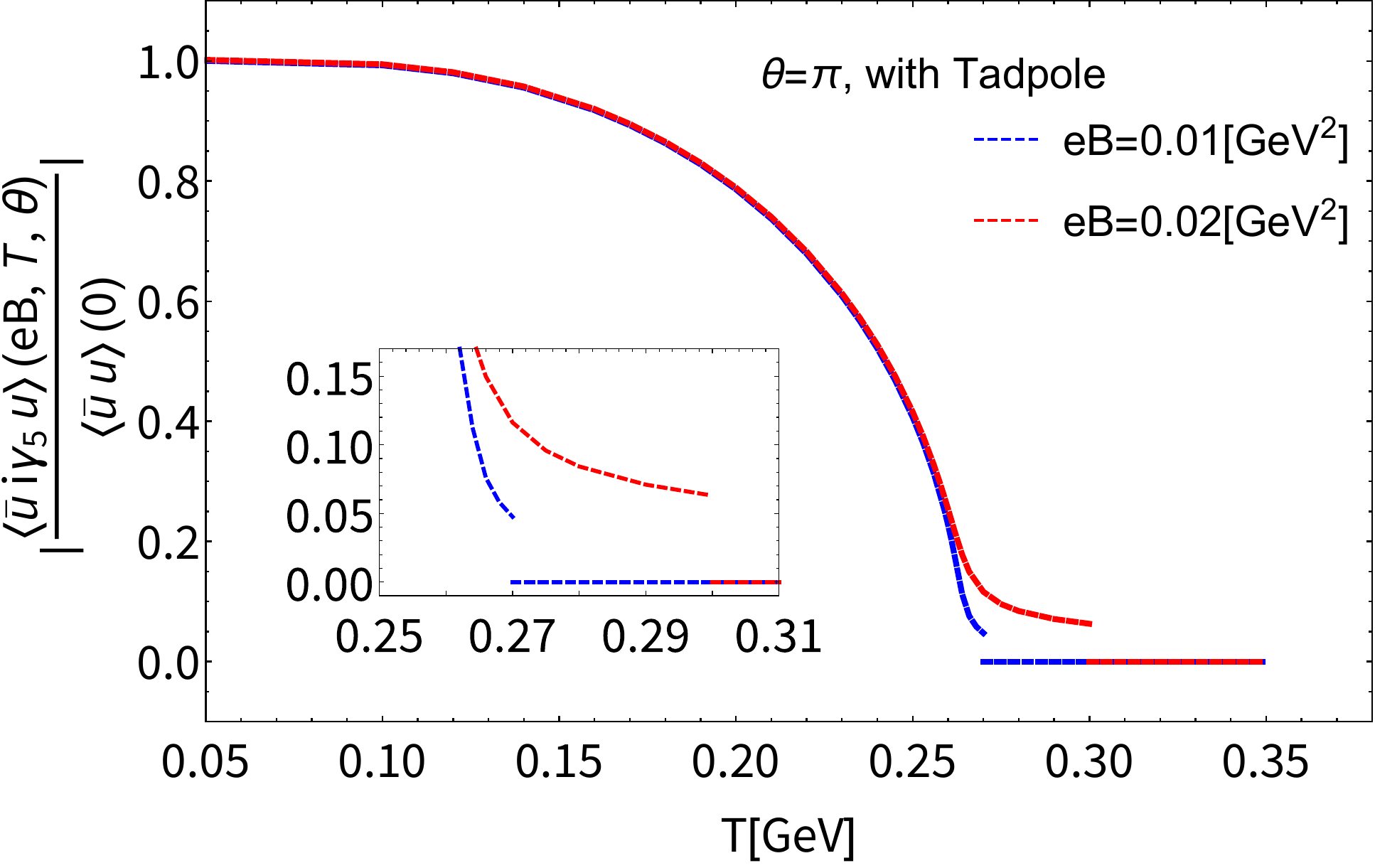}
\caption{Plots of 
$\frac{P_u}{P_u(T=eB=0)} = \frac{\left\langle\bar{u} i \gamma_5 u\right\rangle}{\left\langle\bar{u} i \gamma_5 u\right\rangle(T=e B=0)}$ versus temperature for varying magnetic fields, 
at $\theta =\pi$. Left panel shows the case without the tadpole potential term in Eq.(\ref{Tad}) with Eq.(\ref{tr-anom}), while 
the right panel with the tadpole contribution.   
} 
\label{conden}
\end{figure}

In Fig.~\ref{conden}, we plot the $T$ dependence of $P_u=\left\langle\bar{u} i \gamma_5 u\right\rangle$ normalized to its vacuum value at $T=e B=0$ for various magnetic field strengths, with (right panel) and without (left panel) the tadpole potential contribution. 
Since the down-sector part $P_d$ shows qualitatively the same behavior against $T$ 
as what $P_u$ does, we will not explicitly display it.

In the case with $eB\neq 0$, but without the tadpole contributions (see the left panel in Fig.~\ref{conden}), 
the order parameter for both the chiral and CP symmetries, $P$, undergoes   
the normal second-order phase transition at the criticality (around $\sim 300$ MeV). 
This phase transition is fairly insensitive to the presence of the weak $eB$ (See comparison with the curve for $eB=0$)~\footnote{
The critical point of the CP phase transition at $eB=0$ observed in the current analysis is slightly different 
from what has been seen in the literature~\cite{Sakai:2011gs,Huang:2024nbd} based on \cred{a similar NJL model description} with the same set of the input parameters. 
This is because of the different regularization scheme: the smooth cutoff regulator is used as done in Eq.(\ref{reg}) or not.}, 
although the overall size of $P$ gets slightly amplified by $eB\neq 0$ due to 
the magnetic catalysis for the chiral symmetry breaking, as has been observed in the literature~\cite{Chatterjee:2014csa,Bandyopadhyay:2019pml}.

\begin{figure}[t] 
\centering 
\includegraphics[width=0.65\textwidth]{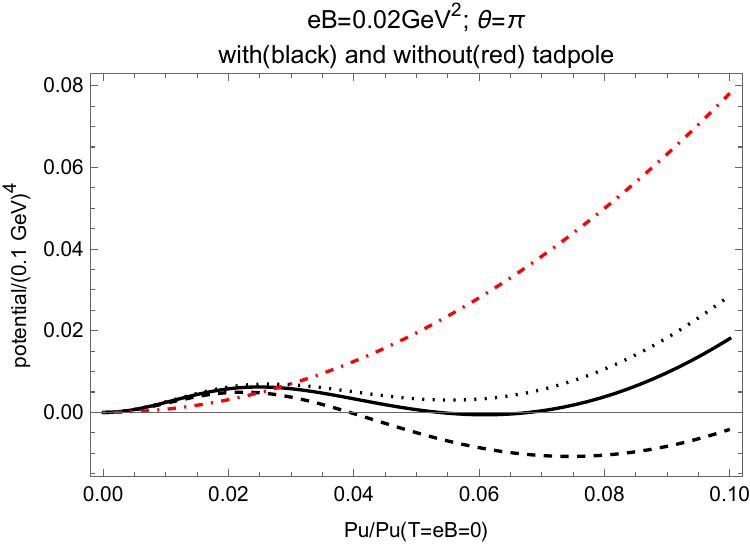}
\caption{The deformation of the thermomagnetic potential 
projected onto 
the $\alpha \propto P$ direction, 
\cred{for the case with the tadpole potential term (black curves), in comparison with the case without 
the tadpole term (red-dashed curve at $T=320$ MeV), 
at $\theta=\pi$, $eB=0.02\,{\rm GeV}^2$, for various values of $T$ around the 
criticality ($T=T_c \simeq 320$ MeV). 
The dashed, solid, dotted black curves correspond to the cases with 
$T=300$ MeV, $320$ MeV, and $330$ MeV, respectively.} 
For simplicity, we have taken the isospin symmetric limit, where 
$P'_u=P'_d=0$ (i.e. $\beta_u=\beta_d=\beta=m_0$), $S'_u=S'_d=S'= - \alpha/(2(g_s + g_d))$, $q_u=q_d=(2/3)e$. 
Potential values have been adjusted by dividing by a factor of $(0.1\,{\rm GeV})^4$. 
} 
\label{Vtot-vs-wo-tad}
\end{figure}

Interestingly enough, 
with the tadpole term contribution (the right panel of Fig.~\ref{conden}), 
the CP phase transition becomes first order and
the criticality is shifted towards higher temperatures as $eB$ increases. 
This first order nature emerges due to the presence of 
the potential barrier generated by the thermomagnetic correction part in the tadpole term, Eq.(\ref{Tad}) with Eq.(\ref{tr-anom}). 
Figure~\ref{Vtot-vs-wo-tad} visualizes the tadpole term 
contribution to the total potential projected along the $P$-direction, 
where we have chosen $eB=0.02\,{\rm GeV}^2$, various temperatures around the criticality. 
For simplicity, we have taken the isospin symmetric limit: 
$P'_u=P'_d=0$, i.e. $\beta_u=\beta_d=\beta=m_0$, 
$S'_u=S'_d=S'= - \alpha/(2(g_s + g_d))$, $q_u=q_d=(2/3)e$. 
One can see from the figure that the tadpole contribution generates 
the barrier around the origin of the $P$-direction.

\begin{figure}[t] 
\centering 
\includegraphics[width=0.48\textwidth]{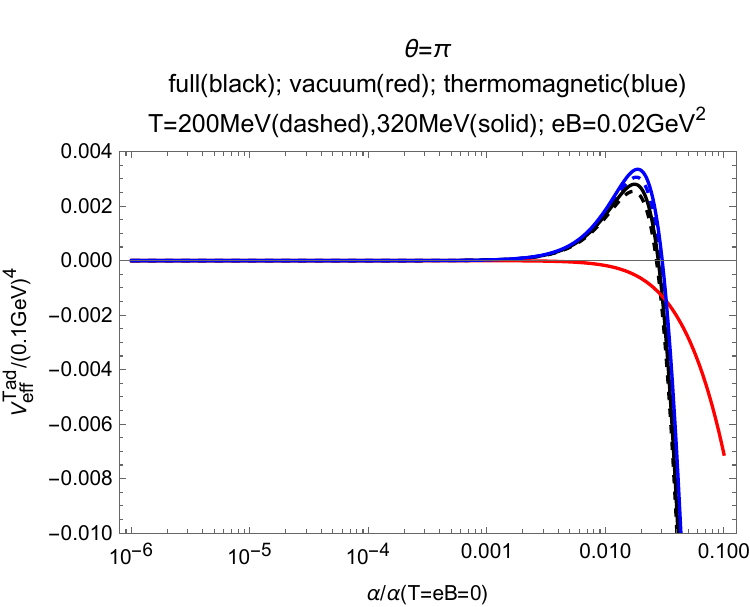}
\includegraphics[width=0.48\textwidth]{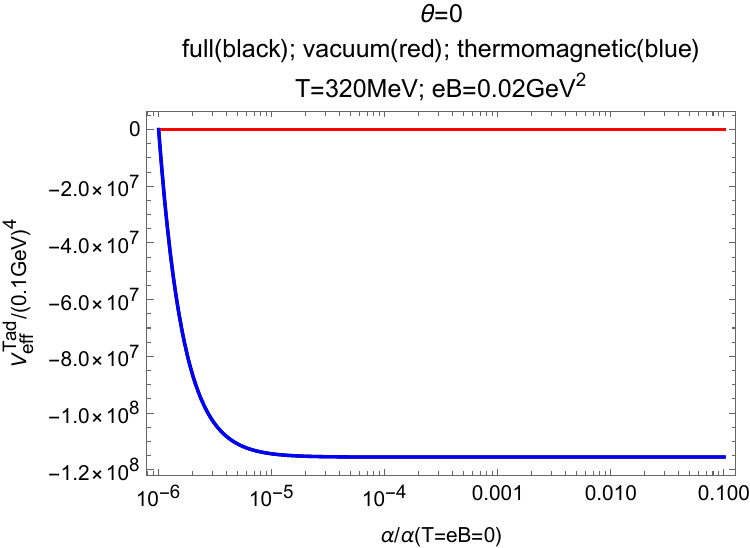}
\caption{(Left panel): the breakdown of the tadpole-potential term-contributions in Eq.(\ref{Tad}), as a function of the CP order parameter $\alpha(\theta=\pi) = - 2P(g_s+g_d)$ normalized by the vacuum value at $T=eB=0$. 
The plots are split into the full contribution (black curve), the vacuum part at $T=0$ (red curve), and the thermomagnetic part (blue curve), for $T=200$ MeV and 320 MeV, with $eB=0.02\,{\rm GeV}^2$ fixed. 
The isospin symmetric limit has been applied as described in the main text. 
(Right panel): the same for $\alpha(\theta=0) = - 2S(g_s+g_d) $. 
The potential values in both panels have been adjusted by multiplying by some factor 
as in Fig.~\ref{Vtot-vs-wo-tad}. } 
\label{Vtad-alpha-plot}
\end{figure}

To clarify the barrier generation mechanism, 
in the left panel of Fig.~\ref{Vtad-alpha-plot} we plot the tadpole term 
contributions at $\theta=\pi$ 
as a function of $\alpha_u$ in Eq.(\ref{alpha-beta}), for $T=200$ MeV and 300 MeV 
with $eB=0.02\,{\rm GeV}^2$ fixed. 
The potential has been normalized as $V_{\rm eff}^{\rm Tad}(0)=0$. 
Again, for simplicity, we have taken the isospin symmetric limit as above.  
A spike structure is observed indeed in the thermomagnetic part (blue curves), 
which is not substantially affected even in the total tadpole term (black curves) 
including the vacuum term at $T=0$ (red curve). 
This sort of a peak structure is not seen when $\theta=0$ along 
the $\alpha$ direction (right panel of Fig.~\ref{Vtad-alpha-plot}), 
where the isospin-symmetric limit gives 
$P_u=P_d=0$, i.e., $\beta_u=\beta_d=\beta=0$, 
and $S_u=S_d=S= - (\alpha - m_0)/(2(g_s + g_d))$.

This discrepancy can be understood in an analytic way. 
First of all, consider only the thermomagnetic term in Eq.(\ref{Tad}) (i.e., the 
second term in Eq.(\ref{tr-anom})) at $\theta=\pi$, and take 
the isospin symmetric limit as above. 
Then the corresponding tadpole potential term goes like~\footnote{
\cred{With things written in terms of $\alpha$ as in Eq.(\ref{V-barrier}), 
the tadpole potential term is free from the model parameter dependence, except $m_0$.}  
} 
\begin{align}
    V_{\rm eff}^{\rm (Tad)} (\alpha) 
    \Bigg|_{\theta=\pi} \approx 
    \frac{|eB|^3}{f_0} \frac{(2/3)^2}{8\pi^2} \sum_{n=0}^\infty 
    \alpha_n \frac{|\alpha|}{{\bf M}_n^2(\alpha)} 
    {\cal I}(T/{\bf M}_n(\alpha)) 
\,, \label{V-barrier}
\end{align}
up to a factor of $(N_f\cdot N_c)=2\cdot 3$, where $\alpha$ at $\theta=\pi$ is related to $P$ in the original base as $\alpha= - 2(g_s+g_d) P$, and 
\begin{align}
    {\bf M}^2_n(\alpha) &=  \alpha^2 + \frac{4 n}{3} |eB| + m_0^2 
\,, \notag\\ 
{\cal I}(T/{\bf M}_n(\alpha)) &=  \int_0^\infty \frac{dz}{(z^2+ 1)^{5/2}} \frac{1}{1 + e^{\frac{\sqrt{z^2+1}}{T/{\bf M}_n(\alpha)}}} 
\,. 
\end{align}
It is manifest from Eq.(\ref{V-barrier}) that 
this thermomagnetic contribution becomes significant at higher $T$ ($T \gtrsim {\bf M}_n(\alpha)$): otherwise 
Boltzmann suppressed. 
At around the criticality $T\sim 300$ MeV, 
the dynamical quark mass $(\alpha)$ has already dropped down to be smaller than 
the current quark mass $m_0$ (see Fig.~\ref{conden}), at which moment 
with $eB = {\cal O}(f_\pi^2)$, only the lowest Landau level essentially contributes. 
Thereby, Eq.(\ref{V-barrier}) can further be approximated to be 
\begin{align}
    V_{\rm eff}^{\rm (Tad)} (\alpha) 
    \Bigg|_{\theta=\pi}^{T\sim 300\, {\rm MeV}; eB = {\cal O}(f_\pi^2)} \approx 
    \frac{|eB|^3}{f_0} \frac{(2/3)^2}{8\pi^2}  
     \frac{|\alpha|}{\alpha^2 +m_0^2} 
    {\cal I}(T/{\bf M}_0(\alpha)) 
\,.  \label{V-barrier:v2}
\end{align}
Now the integral ${\cal I}_\pi$ is almost constant in $\alpha$, so that this tadpole term scales with $\alpha$ and $eB$ as 
\begin{align}
    V_{\rm eff}^{\rm (Tad)} (\alpha)  
    \Bigg|_{\theta=\pi}^{T\sim 300\, {\rm MeV}; eB = {\cal O}(f_\pi^2)} 
    \propto 
    \frac{|eB|^3}{f_0} 
     \frac{|\alpha|}{\alpha^2 +m_0^2} 
\,.  \label{V-barrier:v3}
\end{align}
This has a peak at $\alpha = m_0$, hence a potential barrier is built there, 
and its magnitude gets greater as $eB$ increases. 
\cred{Note also that Eq.(\ref{V-barrier:v3}) is ultraviolet free, hence the emergence of 
the barrier, i.e., the induced first-order feature is free from 
the regularization scheme}~\footnote{
\cred{
In the literature~\cite{Avancini:2019wed}, it has been discussed that in the case of  
the NJL model similar to the present one, 
at $T=0$ an $eB$-dependent regulator as in Eq.(\ref{reg}) leads to an artificial wavy or oscillating behavior of the quark condensate in a strong $eB$ regime ($eB \gtrsim 0.2\, {\rm GeV}^2$). We have checked that for the weak $eB$ regime that we are currently concerned about, 
the CP order parameter $P$ at $T=0$ does not exhibit such a wavy trend.   
}
}.

The peak structure is generated when the method of reducing the quark-thermal loop integral from three dimensions to just one dimension (along the $p_z$ direction) works well. 
Recall that as in Eq.(\ref{replacement}), the spatial momentum integral in Eq.(\ref{V-barrier}) has been replaced by the area related to the magnetic field ($eB$) and a sum over Landau levels. Since near the critical point, the lowest Landau level is dominant in the tadpole term, the $\alpha$ dependence arises in Eq.(\ref{V-barrier}) mainly by the scaling of the one-dimensional loop momentum $p_z$. If the scaling of the other spatial momenta ($p_x, p_y$) or alternatively higher Landau levels were considered, it would introduce additional $\alpha$-dependence, potentially disrupting the barrier generation structure.

When this tadpole term is combined with the non-anomalous potential part ($\Omega$ in Eq.(\ref{Omega-tot})), 
higher barrier tends to shift the criticality to be higher $T$. 
This trend is induced due to the cubic power scaling $\sim |eB|^3$ in Eq.(\ref{V-barrier:v3}). 
This magnetic catalysis has indeed been observed in the right panel of Fig.~\ref{conden}. 
Of particular interest is also to note that the $\alpha$ dependence in Eq.(\ref{V-barrier:v3})
is totally nonperturbative: beyond a simple-minded Ginzburg-Landau description.

On the other hand, 
at $\theta=0$, 
the thermomagnetic part in Eq.(\ref{Tad}) is evaluated as 
\begin{align}
    V_{\rm eff}^{\rm (Tad)} (\alpha) 
    \Bigg|_{\theta=0} \approx 
    \frac{|eB|^3}{f_0} \frac{(2/3)^2}{8\pi^2} \sum_{n=0}^\infty 
    \alpha_n \frac{|\alpha - m_0|}{{\bf M'}_n^2(\alpha)} 
    {\cal I}(T/{\bf M'}_n(\alpha)) 
\,. \label{V-barrier:0}
\end{align}
In this case $\alpha$ at $\theta=0$ is related to $S$ in the original base as $\alpha= - 2(g_s+g_d) S$, and 
\begin{align}
    {\bf M'}^2_n(\alpha) &=  \alpha^2 + \frac{4 n}{3} |eB| 
\,. 
\end{align}
Similarly to Eq.(\ref{V-barrier:v2}), 
Eq.(\ref{V-barrier:0}) can further be approximated at $T \sim 300$ MeV to be 
\begin{align}
    V_{\rm eff}^{\rm (Tad)} (\alpha) 
    \Bigg|_{\theta=0}^{T\sim 300\, {\rm MeV}; eB = {\cal O}(f_\pi^2)} 
    &\approx 
    \frac{|eB|^3}{f_0} \frac{(2/3)^2}{8\pi^2}  
     \frac{|\alpha - m_0|}{\alpha^2} 
    {\cal I}(T/\alpha) 
    \notag\\ 
    & \propto 
    \frac{|eB|^3}{f_0} 
     \frac{|\alpha - m_0|}{\alpha^2} 
\,.  \label{V-barrier:0v2}
\end{align}
This does not develop any peak in contrast to the $\theta=\pi$ case.  
Note that both Eq.(\ref{V-barrier:v3}) and Eq.(\ref{V-barrier:0v2}) become identical 
when $m_0 =0$. In other words, the discrepancy between two cases (with $\theta=0$ or $\pi$) has been generated by nonzero current quark mass.

Other remarks are in order: 
\begin{itemize}
    \item  
the electromagnetic interaction 
does not break the CP symmetry, hence does not destabilize 
the symmetric phase $\alpha(\theta=\pi) \propto P=0$, which is reflected in 
the potential form in Eq.(\ref{V-barrier:v3}), while 
does explicitly break 
the chiral (and $U(1)$ axial) symmetry, therefore, the chiral symmetric phase $\alpha(\theta=0) \propto S=0$ 
is no longer stabilized, as in Eq.(\ref{V-barrier:0v2});  

\item 

in the chiral limit $m_0 \to 0$, the $\theta$ parameter becomes unphysical, so 
the CP order parameter $P$ coupled to $\theta$ as well as 
the CP phase transition is ill-defined unless another CP violating parameter is introduced. Therefore, the discrepancy between two cases (with $\theta=0$ or $\pi$) 
should be seen as long as quarks are massive only when the $\theta$ parameter is physical.

\end{itemize}

\section{Summary and discussion}

In summary, 
we have discussed the thermal CP phase transition in QCD at $\theta=\pi$ under a weak magnetic field background with $eB \lesssim \cred{\Lambda_{\rm QCD}^2}$, where the electromagnetic scale anomaly plays a significant role. 
We have worked on a two-flavor \cred{NJL} model at $\theta=\pi$ in the MFA, including the electromagnetic-scale anomaly term (Eqs.(\ref{Tad}) with (\ref{tr-anom})). 
The thermal CP phase transition has been observed to be of first order, in sharp contrast to the two-flavor (P)NJL without the electromagnetic scale anomaly (Figs.~\ref{conden} and~\ref{Vtot-vs-wo-tad}).

The electromagnetic-scale anomaly effect is dominated by 
the thermomagnetic term (the second term in Eq.(\ref{tr-anom})),
which significantly creates the potential barrier at high temperatures  
around the origin of 
the CP-order parameter ($P$) direction (Fig.~\ref{Vtad-alpha-plot}). 
This barrier generation is induced by non-standard potential or nonperturbative form $\sim |eB|^3 \frac{|P|}{P^2 + m_0^2}$ (Eq.(\ref{V-barrier:v3})), in contrast to 
the CP-invariant chiral phase transition at $\theta=0$ (Eq.(\ref{V-barrier:0v2})), 
although both forms are reduced to be identical to each other in the chiral limit $m_0\to 0$. 
This implies a significant amplification of the $U(1)$ axial violation 
due to the electromagnetic scale anomaly at high temperatures when $\theta=\pi$.

The present analysis has been based on the model of conventional NJL type, 
with which the inverse magnetic catalysis is not realized in a strong $eB$ regime 
at $\theta=0$, in contradiction to the lattice result~\cite{Bali:2011qj}. 
The electromagnetic scale anomaly survives only in a weak $eB$ regime, like 
$\sqrt{eB} \lesssim 100$ MeV, 
while the current lower bound of the magnetic field strength $\sqrt{eB}$ on lattices 
is $\sim$ 100 MeV~\cite{Bali:2011qj}. 
The inverse magnetic catalysis has therefore not yet been observed in such a weak $eB$ regime. Nevertheless, it would be intriguing to incorporate 
an $eB$-dependence into the scalar-four fermion coupling ($g_s$) to be fitted to the 
lattice data on the $eB$-dependence reflecting the inverse magnetic catalysis of the 
chiral crossover, as has been done in the literature even at $\theta=\pi$~\cite{
Chatterjee:2014csa,Bandyopadhyay:2019pml,Carlomagno:2025ayh}, to observe how the first-order nature of the CP phase transition could be weaker or still stay strong enough against the inverse magnetic catalysis. This is to be explored elsewhere.

An extended Columbia plot at $\theta=\pi$ with varying quark masses under 
a weak magnetic field background can also straightforwardly be investigated in a way 
similar to the work in~\cite{Kawaguchi:2021nsa} for the chiral phase transition. 
In the present analysis, we have also found that the potential barrier generation thermomagneticly induced from the 
electromagnetic scale-anomaly seems to persist 
as long as the up and down quark masses are in a range 
$10^{-5} < m_0/m_{\rm phys} < 10 $, where 
$m_{\rm phys}$ denotes the value at the physical point. 
In the case below the lower bound, the CP-even and -odd chiral-order parameters 
$S$ and $P$ behave like almost identical $(V_{\rm eff}^{\rm (Tad)} \propto 1/|\alpha|)$ in the tadpole term contribution (see Eqs.(\ref{V-barrier:0v2}) and (\ref{V-barrier:v3})), hence the intrinsic barrier structure 
is gone in the $P$-direction. When the quark mass exceeds the upper bound, 
no spike structure or nontrivial contribution in the $P$-direction will be left. 
The parameter space in the two-flavor NJL model outside the first-order regime above would all be the second order regime, 
unless another explicit CP violation other than what is induced from nonzero $\theta$ is introduced. 

\cred{Inclusion of the Polyakov loop variable into the NJL model 
could give a nontrivial result on the deconfinement phase transition at $\theta=\pi$: 
the tadpole potential term induced from the electromagnetic scale anomaly would get 
the Polyakov loop dependence via the Fermi-Dirac distribution function part in $F$ in the 
thermomagnetic term (see Eq.(\ref{F})). Note that the precise form of the Fermi-Dirac distribution function part in $F$ will not substantially be sensitive to the generation of 
the barrier around the criticality, which acts like almost constant (see Eqs.~\ref{V-barrier:v2} and~\ref{V-barrier:v3}.), so that the first order nature will still survive with 
the Polyakov loop extension. 
Consequently, 
the thermal evolution of the Polyakov loop variable,
would be nontrivially affected by the barrier generation mechanism along the CP order parameter, 
due to the modification of the peak structure as demonstrated in Eqs.(\ref{V-barrier}) - (\ref{V-barrier:v3}). 
The deconfinement phase transition might thus become 
first-order like. 
Actually, this extension involves a theoretically nontrivial issue on 
the correlation between the scale anomaly and confinement, and would provide more phenomenological consequences other than the CP phase transition. 
This issue is thus to be left to be pursued for another publication. 
}

\cred{When the strange quark is incorporated into the model, 
the intrinsic potential barrier structure at $T=0$ associated with 
the cubic potential term arising from the $U(1)$ axial anomaly. 
However, the electromagnetic scale-anomaly induced barrier is created near the origin ($P\sim 0$), while the cubic term becomes significant for larger $P$. 
Therefore, the emergence of the former barrier structure is expected to persist even in the three-flavor case. }
Detailed study on this issue deserves another publication.

At any rate, the presently observed first-order nature of the CP phase transition 
will provide a new benchmark toward understanding of 
QCD at $\theta=\pi$, prior to the future development in lattice simulations 
and the 't Hooft anomaly matching condition. 
Though being effective enough to make such a first step, 
the current model analysis can be improved 
by going beyond the MFA, say, using 
the functional renormalization group method.

In closing, we give comments on the phenomenological and cosmological applications. 
The first-order criticality of the CP phase transition in a weak magnetic background 
could give an impact on the interpretation of the observed nano-hertz gravitational-wave in terms of axionlike particle-domain wall collapses, as was addressed in~\cite{Huang:2024nbd}, and/or might also be relevant to 
the generation of primordial black hole.   
Those are also noteworthy to be pursued in another publication.

\section*{Acknowledgments} 
This work was supported in part by the National Science Foundation of China (NSFC) under Grant No.11747308, 11975108, 12047569, 
and the Seeds Funding of Jilin University (S.M.). 
The work by M.K. is supported by RFIS-NSFC under Grant No. W2433019.
The work of A.T. was partially supported by JSPS  KAKENHI Grant Numbers 20K14479, 22K03539, 22H05112, and 22H05111, and MEXT as ``Program for Promoting Researches on the Supercomputer Fugaku'' (Simulation for basic science: approaching the new quantum era; Grant Number JPMXP1020230411, and 
Search for physics beyond the standard model using large-scale lattice QCD simulation and development of AI technology toward next-generation lattice QCD; Grant Number JPMXP1020230409).

\end{document}